\documentclass{article}

\title{The AGN tori sizes: a remark on astro-ph/0512025 by Moshe Elitzur}
\author{Gian Luigi Granato \& Luigi Danese}

\begin{document}

\maketitle

\begin{abstract}
\noindent We point out that in Granato \& Danese 1994 and Granato
et al.\ 1997 we predicted maximum observable sizes for the
putative torus in NGC1068 of 10-20 pc, not "hundreds of parsecs"
as stated by M.\ Elitzur in astro-ph/0512025.
\end{abstract}

In astro-ph/0512025 Moshe Elitzur persists with misleading
statements on the results of our radiative transfer models for
tori around AGN (in particular Granato, Danese \& Franceschini,
1997, ApJ, 486 147, but see also Granato \& Danese 1994, MNRAS,
268, 235 and Galliano et al.\, 2003, A\&A, 412, 615). Namely, the
reader gets the wrong message that we predicted sizes of hundreds
of pc for objects such as NC1068 (and similar or even less
luminous objects), while now high resolution mapping in the NIR
and MIR give upper limits to the size of the torus of the order of
a few tens of pc. This interpretation of our predictions is by far
erroneous, and may lead to misuse of our results. The general
conclusion of our modelling efforts has been instead that the
ratio between the outer radius and the dust sublimation radius
$r_{out}/r_{in}$ should be of the order of 100-300 for typical
Seyfert 1 and 2 galaxies, where the dust sublimation radius
$r_{in}$ scales with the square root of the intrinsic AGN
luminosity. Expressing the luminosity in units of $10^{46}$ erg/s,
as a rough rule of thumb we have $r_{in} \sim 0.5 L_{46}^{1/2}$.
Then AGN whose luminosity is comparable to that of NGC1068
($L_{46}\sim 0.06$), should have tori much less extended than 100
pc.

Specifically, in Granato et al (1997) we reproduced the SED of
NGC1068 with a torus having an external radius of 19 pc
($r_{out}/r_{in}=150$), and this was the largest value we ever
used (previously in 1994 we used 12 pc for NGC1068). We refer the
reader to Figure 1 in Granato et al.\ 1997, keeping in mind that
the radius of 30 pc reported in that caption, already much smaller
than "hundreds" pc, has to be rescaled by a factor 22/14.5,
because at those times we adopted a distance for NGC1068 of 22 Mpc
rather than the now standard value of 14.5. Another point to
consider is that 19 pc is the size of the whole model torus, while
observations at shorter and shorter wavelength picks out inner and
inner regions, where the dust is hotter and hotter. To have a
qualitative feeling of this, please have a look to Figure 5 of the
same paper (the precise $\lambda$ behavior depends on the adopted
PSF and on details of the geometry that, as discussed in Galliano
et al.\ 2003, cannot be inferred from SED fitting alone. Also, in
this paper we made the very clear point that even the size of the
torus, one of the best constrained geometrical parameters by SED
fitting, it is still uncertain by a factor at $\sim 2$).

In general, we always used in our published SED fittings of NGC
1068 tori with a ratio between the outer and inner (sublimation)
radius between 30 and 150, ratios that translate into outer radii
of the torus between 4 and 20 pc. However, only observations at
$\lambda > 20 \mu$m would not be dominated by regions
substantially smaller than this.

Thus, our work should not be improperly involved in the origin of
idea of $>100$ pc scale tori for AGNs with luminosity comparable
to that of NGC1068 (which actually seems a sort of 'media' bluff).
If this has been done, it is due to superficial or biased reading
of our papers, and possibly to optimistic writings of proposals
for high resolution imaging, not an our fault (it is conceivable
that people wanted to write that we predicted easy to see tori).
Instead, we should be quoted as supporters of tori whose
$r_{out}/r_{in}$ is of the order of 100 rather than 3-10, as
assumed by Pier and Krolik in their seminal works, well justified
by the scanty information available at that time. Finally, we
point out that the original Pier and Krolik (1993, ApJ, 418, 673)
models would imply a TOTAL radius of the NGC1068 torus $<1pc$.

\end{document}